Chapter 3

# Control and Energy Management System in Microgrids

*Hajir Pourbabak, Tao Chen, Bowen Zhang and Wencong Su*

## 3.1  Introduction

The U.S. Department of Energy defines a microgrid [1] as "a group of interconnected loads and distributed energy resources (DER) within clearly defined electrical boundaries that act as a single controllable entity with respect to the grid. A microgrid can connect and disconnect from the grid to enable it to operate in both grid connected and island mode." It is interesting to mention that the concept of a microgrid has been around for more than a hundred years. Around 1880, Thomas Edison founded and established the first investor-owned electric utility on Pearl Street in lower Manhattan of New York City [2]. On September 4th, 1882, the Pearl Street power station went into operation. This small electric utility was allowed to operate its 27-ton "Jumbo" constant-voltage dynamo (steam generator) and serve 82 local customers without being connected to a main grid, which did not exist yet. This investor-owned electric utility can be considered as the very first version of the microgrid, as shown in Figure 3.1.

In the early 1900s, the first statewide regulation of electric utilities emerged [3]. Due to the evolution of interconnected power grids through long transmission lines, the electric utilities were moving from a microgrid-like independent system to a highly centralized and regulated one. Across the world, the development of the microgrid had been fairly silent until the early 2000s. In the past two decades, however, the original microgrid concept has drawn increased attentions to address the limited electricity access issues for in remote and less developed communities. Microgrids are often the only practically possible solution or the most cost-effective way for these areas that are not connected with to the utility grid. In addition, the enhanced microgrid concept offers new socio-economic benefits that have not even been imagined previously. For instance, non-traditional power generators (e.g., wind turbines, solar panels, small-scale diesel generators) in microgrids are allowed to sell electricity to local consumers, ultimately boosting electricity market restricting activities. In addition, the microgrid no longer relies on a single power source; the on-site generation can be used as an emergency backup in the event of a blackout or brownout in order to mitigate the disturbance and improve power reliability.

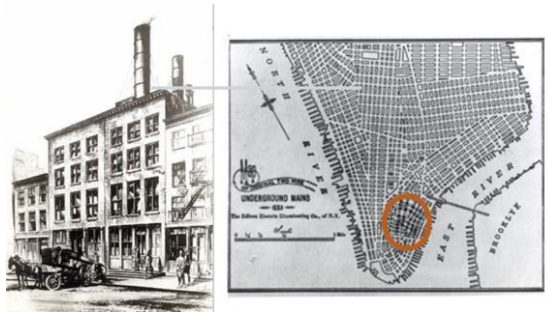

Figure 3.1: Edison strategically located the station in the densely populated area of lower Manhattan. (Photo credit: ConEd)

Figure 3.2 demonstrates the concept of modern microgrids. Technically speaking, a modern microgrid is a small portion of a low-voltage distribution network that is located downstream from a distribution substation through a point of common coupling (PCC) [4]. Due to the nature of microgrid operations (e.g., ownership, reliability requirement, locations), a major microgrid deployment is expected to be carried out on university campuses and research institutions, military bases, and industrial and commercial facilities. According to Navigant Research (formerly called Pike Research), the global microgrid capacity is expected to grow from 1.4 GW in 2015 to 7.6 GW in 2024 under a base scenario [5]. Modern microgrids not only offer great promise due to their significant benefits, but also result in tremendous technical challenges. There is an urgent need to investigate the sophisticated and state-of-the-art control and energy management systems in microgrids.

The remainder of this chapter is organized as follows: Section 3.2 discusses the protection and control aspects of a microgrid. Section 3.3 discusses the energy management aspect of a microgrid. Section 3.4 introduces the demand response and demand side management. Section 3.5 briefly discusses the home energy management systems. Section 3.6 presents the energy management system with Supervisory Control And Data Acquisition (SCADA) system. Section 3.7 covers the supporting infrastructure of a microgrid, including smart meters, advanced metering infrastructure (AMI), and communication infrastructure. Section 3.8 summarizes the major contributions of this chapter and briefly discusses the future research trends.

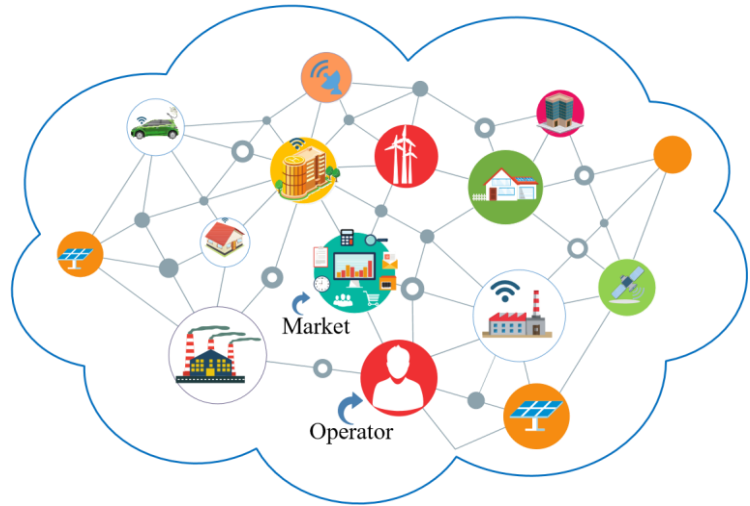

Figure 3.2: An illustrated microgrid system architecture

## 3.2   Protection and Control of Microgrid

The electrical energy generated by wind farms, solar energy and even small local generators inside of the microgrids is reaching a considerable portion of the total produced energy in comparison to that of the previous decade. The presence of new energy sources, distributed storage, power electronic devices and communication links make a power system's control and protection more complicated than before [6] because they impose considerable and fundamental changes in the configuration of the power system topology and power flow direction [7]. Thus, in order to enhance the power system visibility and controllability, more data and communication links should be provided throughout the entire power system; however, this huge amount of data can also cause heavy computational burden, as well as possibly negatively impact the performance of protection schemes.

As mentioned several times, in recent decades, more attention has been given to the microgrids framework from the aspects of the market, control, management, reliability, etc. due to the active role of both the energy producers and consumers. A microgrids that could be a kind of smart grid provides us with more flexibility and reliability for control and protection of a power system. Live interaction between private commercial generators and controllable consumers is an inseparable part of a smart grids that makes the power system more and more complex to handle. Thus, it is admittedly evident that conventional protection and control systems will not effectively work in a microgrid because they cannot satisfy all the control and protection requirements of such a dynamic and variable grid. The importance of the inescapable integration of communication and the physical energy network (i.e. the power system) needs to be taken into account as a way to reach an advanced and developed management system for a future grid-connected microgrids [8, 9, 10].

### 3.2.1 Microgrids Protection

Microgrids protection and reliability are the most serious challenges in the area of power system protection due to the complex nature of microgrids, two-way flow of both power and information in the power system, and presence of local distributed generations. One of the protection issues microgrids suffer from is islanding. Islanding will happen if a microgrid that includes distributed generation/storage and local loads becomes separated from the rest of the power grid. The island area brings its own problems, such as low power quality, safety matters, and overload issues [11].

Islanding detection methods are an immediate and effective solution for failure, disconnection and outage inside microgrids. Researchers have done great efforts to both introduce new detection methods and advance previous ones [12, 13, 14, 15] to lessen the adverse effects of the islanding phenomenon. However, the smart technology applied to future microgrids provides effective tools and fundamentals to use islanding detection as a last resort to avoid any consequent trouble. As Fang *et al.* [16] classified, two protection mechanisms can be named: fault prevention and fault detection/recovery. The former mechanism is normally suitable as a preventive measure for avoiding any failure as much as possible, while the latter is used after failure to detect and remove the fault and recover the system in the shortest possible amount of time. Several approach are proposed to improve the fault detection and recovery process. Tate *et al.* [17] introduced an algorithm that uses measured data of PMU to successfully detect line outages based on information regarding the system topology. It is mentioned in [18] that phasor measurement for detection of system parameters' error is necessary because conventional measurement cannot provide a powerful tool to identify such certain error.

One of the approaches for enhancing the system reliability is self-healing ability. This self-healing ability helps the system to reconfigure itself based on the intended pattern and partitions the system into some small intentional island. This reconfiguration will be done against the disturbances that cannot be removed. [19] proposed a controlled area partitioning algorithm to break down the whole system under danger into some island to minimum active and reactive power imbalances. The most important contribution of the proposed method is its improvement of the voltage profile of the partitioned subsystems compared with those of similar algorithms. In fact, cascading failure will be prevented and the impact of all disturbances will be restricted [16].

As previously mentioned, the flow of information through the communication network is important as same as flow of power through transmission line. Thus, any missing data, failure on communication channel or problem within smart meter could cause a serious issues with regard to microgrids operation, visibility and controllability.

### 3.2.2 Control Approach of Microgrid Control

In this chapter, different microgrid control methods ranging from conventional to recently introduced ones are studied and categorized into three major groups: centralized, decentralized and distributed control methods.

In a power system, the control of generators and their economic dispatch can be carried out by a centralized method that much research has been dedicated to this approach [20]. In this approach, data and information are gathered from all over the system and will be processed in a central controller; then, a control signal will be transmitted directly to each agent through supervisory, control, and data acquisition (SCADA). SCADA is an advanced automation control system that centrally manages the control, gathering and monitoring of an electrical power system's operation [21]. As can be seen from Figure 3.3, a two-way communication channel for each agent is required for transferring data to the center of the system, based on the nature of the centralized approach. Therefore, the system will be confronted by a barrier due to the numerous communication links that become necessary as the number of agents increases [22]. Moreover, the centralized method is not an effective or practical approach in the future grids because both the communication and electrical network topologies of a microgrid are always subject to change [23]. All that aside, using the centralized method for a large number of agents may not be a cost-effective solution due to the need for a high level of connectivity. Implementation of the centralized method, although theoretically easy, is still not a straightforward way to expand power systems in the future. However, it is obviously proven that centralized methods have currently become a mature and developed approach for power system and cannot be replaced completely by other new methods. Thus, any new method must be compatible with the current infrastructure of a power system to work successfully with it. Solving an economic dispatch (ED) problem for a large scale power system is a good example of widespread use of the centralized method.

As mentioned before, one type of control method is the decentralized method, in which each agent or subsystem has its own controller, meaning each agent makes decisions based on local measurements, such as voltage and frequency value, to get more profit from the market or more stability. In contrast to the centralized method, which needs global information to make a decision for the whole system, the decentralized method will not require agents to exchange all information with other agents or a central control, neither globally nor locally; however, some leader agents can send and receive information through the center. Due to the absence of communication links, decentralized approach cannot guarantee that the system will reach the global optimization or stability; thus, they are used for local improvement. However, the decentralized method needs to be considered from different angles because each method has its own pros and cons. One of the important advantages of decentralized systems is their ability to protect the agents' privacy by secluding their private information [24, 25, 26]. A centralized and decentralized approach have been provided at [27] to improve power system transient stability in a small scale power system and authors also discuss and compare the performance of both methods.

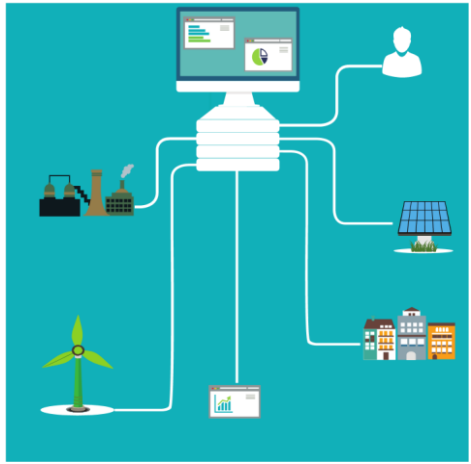

Figure 3.3: Centralized method

Furthermore, a decentralized system is more stable than an identically connected centralized system; for instance, if some leaders lose their connection with other agents, we still have some decentralized systems that can remain stable. A simple representation of the decentralized method is shown in Figure3.4.

Recently, painstaking research has been done to find an alternative method, or at least a short-term solution, to reduce some of the problems of the centralized method. Now, the control of a sophisticated system, such as a power system that uses a distributed control approach for achieving an optimal point, has drawn more and more attention due to its considerable ability to extend a complex system more easily than a conventional central control approach can. In the distributed consensus-based control approach (commonly known as the distributed control approach), each agent (i.e. generators and users of a microgrid) uses local information provided by its neighbors and locally measured parameters, such as voltage and frequency. Figure 3.5 shows the concept of a distributed control system that uses local communication links. In the distributed approach, local agents will share their information among one another through two-way communication links. In other words, in contrast to decentralized systems, in which agents use local measurements without sharing information with their neighbors, distributed systems provide a suitable environment that allows users to share information; thus, the distributed method can achieve global optimization, just like the centralized method.

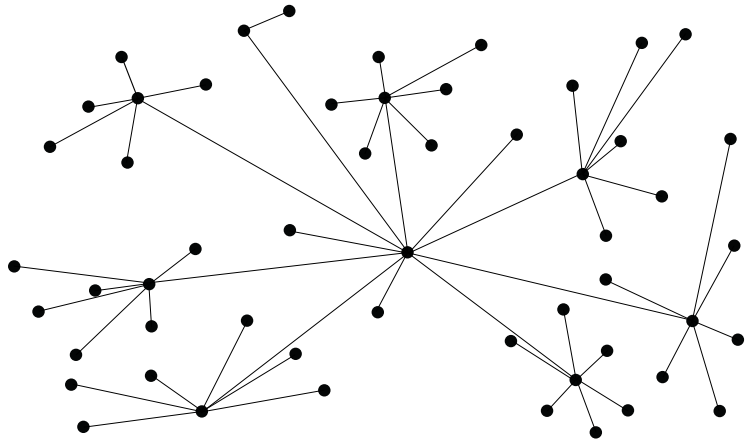

Figure 3.4: Decentralized methods schematic

According to new investigations conducted by different researchers, the distributed approach could be practical for handling the variable nature of the microgrid, which is due to its plug-and-play characteristic. In fact, this method will not be affected by changes to the smart grid topology due to the sharing of local information, and thus would be easy to extend as new agents arbitrarily connect to the network. From figure 3.6, it can be shown how a simple microgrid can be equipped with the distributed method.

Solving economic dispatch problems is a famous and frequent use of the distributed algorithm in microgrids and power systems. In the electrical market, different participants, including consumers and prosumers, have their own private cost function affecting their actions in the market. It is evidently believed that they must consider privacy principals if they want to get more satisfaction and benefits in the electrical competition market. To a great extent, the distributed algorithm can preserve the agents' privacy because most of the important information, at least, will not be released and shared globally.

Olfati-Saber *et al.* [28] wrote a paper providing a theoretical framework for the analysis of a consensus based algorithm applied to a multi-agent system. [29] carried out a comprehensive survey of distributed consensus based problems. The authors concentrate on the application of consensus problems in cooperative control. Deng *et al.* [8] proposed a real-time demand response algorithm to find an economic solution through the participation of both users and producers. They have tried to model a real-time interaction in a smart grid given a multibuyer-multiseller system. Also, this algorithm is promised to be a privacy protector. Rahbari-Asr *et al.* [30] worked on an incremental welfare consensus based algorithm (IWC) among responsive smart load and dynamic DGs inside a smart grid for the purpose of energy management. The IWC algorithm can converge on a global optimum social welfare without having a central controller. Fortunately, the authors are not concerned about scalability, which is verified by a Monte Carlo simulation. Furthermore, this paper states that its proposed algorithm does not need to disclose any private information regarding the utility and cost function. Mudumbai *et al.* [31] represented a distributed algorithm having dual-application. This algorithm can independently control generators' output power in response to frequency deviation, while considering the economic dispatch of generators in a microgrid. Some key features of distributed algorithms have been emphasized for the proposed algorithm, such as scalability, dynamic response, and model independence. The researchers also benchmarked their algorithm's performance against the centralized approach using numerical results.

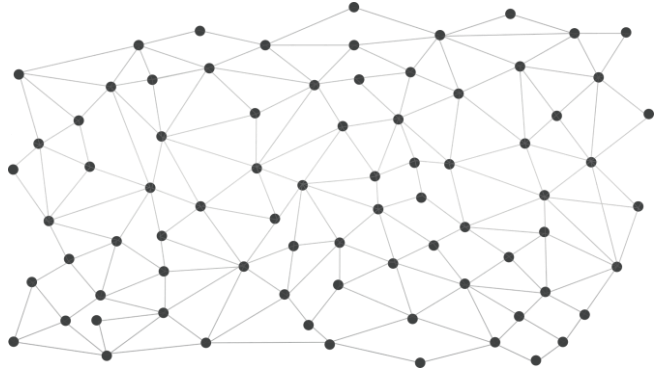

Figure 3.5: Distributed method

Table 3.1 provides a comprehensive and meaningful comparison of the three methods to clearly highlight the pros and cons of each approach [30, 32, 33].

## 3.3 Energy Management Aspects of Microgrid

In the microgrid application scenario, one of the challenging tasks is reducing large energy imbalances due to the uncertainty in power supply from intermittent renewable energy source based distributed generators (DGs) and the dynamic nature of electricity consumption [34]. Fortunately, advances in information and communication technologies (ICT) along with more and more heterogeneous flexible loads, such as plug-in electric vehicles (PEVs), thermostatically controlled loads (TCLs) and distributed energy storage (DES), enable a great opportunity to develop the demand response (DR) and demand side management (DSM) in smart grid applications. These technologies provide a lot of energy management approaches to eusure that the power demand can be rescheduled according to the power supply from utilities or local microgrids through directly or indirectly load control strategy [35]. Within the context of various nondispatchable renewable resources based microgrid, many demand response programs supported by home energy management system (HEMS) or Supervisory Control and Data Acquisition (SCADA) can further promote the participation of active energy customers into power distribution network to provide a way that they can contribute to the optimization of the value chain through directly controlling the self-generated power and electric devices. The potential demand elasticity offered by end-users (e.g. household demand) can postpone or defer grid investments and promote the efficient exploitation of the renewable electricity produced at or close to the consumption level [36]. The implementation of these opportunities require us developing new operational strategies, value mechanism and ICT tools for enabling the coordination between demand scheduling and microgrid with the objective of supporting the entire power distribution network through providing ancillary services. The feasibility of combined optimal operation of microgrids can also be improved by embedding various DR or DSM strategies into the operation.

| Type | Pros | Cons |
| --- | --- | --- |
| Centralized control | <ul><li>Easy to implement</li><li>Easy to maintenance in the case of single point failure</li></ul> | <ul><li>Computational burden</li><li>Not easy to expand (so it is not suitable for smart grids)</li><li>Single point of failure (highly unstable)</li><li>Requires a high level of connectivity</li></ul> |
| Decentralized control | <ul><li>Local information only</li><li>No need for a comprehensive two-way high-speed communication</li><li>Without leaders, system still includes some control island-area</li><li>Parallel computation</li></ul> | <ul><li>Absence of communication links between agents restricts performance</li><li>Moderate scalability</li></ul> |
| Distributed control | <ul><li>Easy to expand (high scalability)</li><li>Low computational cost (parallel computation)</li><li>Avoids single point of failure</li><li>Suitable for large-scale systems</li><li>Not affected by changes in system topology.</li><li>practical solution for plug-and-play characteristic of smart grid</li></ul> | <ul><li>Needs synchronization</li><li>May be time-consuming for local agents to reach consensus</li><li>Convergence rates may be affected by the communication network topology</li><li>Needs a two-way communication infrastructure</li><li>Cost to upgrade on the existing control and communication infrastructure</li></ul> |

Table 3.1: Comparison of Centralized, Decentralized and Distributed Control method

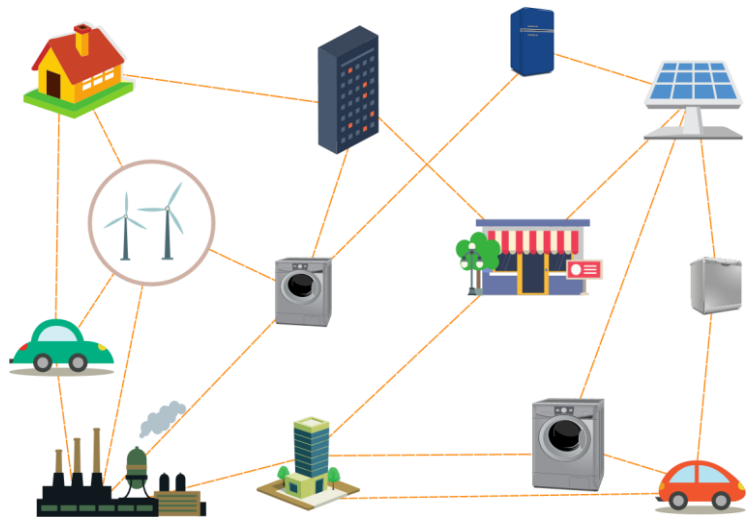

Figure 3.6: Microgrid equipped with distributed approach

Both DR and DSM are mainly aim settling down the energy imbalances caused by irrational energy consumption or optimizing the consumption strategies by aligning the energy consumption to the supply and response immediately to the electricity price signal. Majority of the DR/DSM strategies are designed to reduce the peak demand by shifting the energy demand from peak hours to off-peak hours, namely peak shaving or valley filling.

## 3.4  Demand Response and Demand-side Management

In the last few years, there have been more and more retailers and utilities investing in DR programs, utilizing changes in end-users' electricity demand as one of the ways
10

to increase electricity demand elasticity. Usually, most DR actions may be either responses to changes in the electricity prices over time, or incentives from utilities that result in peak shaving or even the relief of congested networks incentive agreement [37]. With the development of networked microgrids, those incentives also include local power supply situations and relevant generation forecast. Generally, there are two demand response mechanisms, namely incentive-based and price-based. Each DR mechanism comprises a number of DR alternatives that can be adopted, which are shown in Table 3.2.

More specifically, the incentive-based DR mechanism has two subcategories. One of them is the conventional mechanism that is widely used in many applications, including direct control and an interruptible/curtailment program. The other one is innovative market based mechanism that includes emergency actions, demand bidding, reserving market and various kinds of ancillary services markets. On the other hand, there are also several alternatives for the priced based DR mechanism. It includes time-of-use (TOU) pricing, critical peak pricing (CPP), extreme day critical peak pricing, real time pricing and so on. From economic perspective, the benefits of DR actions may be significant for both utilities and customers, if those electricity price mechanism are introduced in a proper way. The cost reduction for both retailers and end-users may be significant reaching up to 18% when 40% of the controllable devices are considered [36]. As the difference between peak and off-peak TOU rates increases, demand elasticity increases as well. These mechanisms can be implemented through the main architecture of Demand Side Management (DSM) framework, which is shown in Figure 3.7. It is noteworthy that DSM techniques depend heavily on two-way communication techniques including wide area network (WAN) and home area network (HAN). The realization of the necessary demand response actions usually requires frequent communication between customers and utilities or local microgrids, especially considering real-time pervasive uncertainty of the highly dynamic intermittent renewable sources, caused by weather conditions.

However, the main barriers for wide rollout of demand response programs as identified by different stakeholders are low consumer interest and ineffective program design. There is also a high correlation between these two barriers because if some more effective program designs were proposed, some of them would possibly encourage customers to actively participate. Otherwise, the benefit will not be big enough to improve customers' interest. So far, the majority of the applied DR mechanisms are based on highly centralized control concepts. Thus they require the acquisition and processing of a very large amount of local information and configuration data from a central point, conflicting with the popular distributed control approaches for microgrids [38]. This exhibits considerable complexity and burden on the only control center, which affects the scalability of aforementioned various DR mechanisms. Most of the manageable demand actions in DR program implementations just concerns large commercial or industrial customers, failing to incorporate a considerable share of small residential customers even with self-generation capability. Thus some low-level energy management systems are needed to deal with energy consumption of residential customers.

| DR program | Time of Use (ToU) | Critical Peak Pricing | Real Time Pricing (RTP) | Direct Load Control | Interruptible | Bidding | Emergency |
|---|---|---|---|---|---|---|---|
| Mechanism type | Price based | Price based | Price based | Incentive based | Incentive based | Incentive based | Incentive based |
| Rule | Non-dispatchable | Both | Non-dispatchable | Dispatchable | Dispatchable | Dispatchable | Dispatchable |
| Response type | Customer side | Customer side | Customer side | Utility side | Customer side | Customer side | Utility side |
| Advantages | Low price rate during off peak, user can shift load with min. | Customer response for a short time period to get discount offers. | The customer can minimize the cost with respect to price change in a day, month, or | The utility offers good discount for limited load reduction or shifting. | Customers respond for a short period to get discount rates. | The utility offers good discount for limited load reduction or shifting. | Customer can get credit or discount rate for the short response. |
| Disadvantages | One price rate for all customers' consumption levels, user should follow the price change with respect to time. | The customer should shift or curtail home resource for certain time. | Customers need to instantaneously respond to minimize bill cost. | The customer should give the utility company a level of authority to shift or curtail certain load in order to balance | The customer should shift or curtail home resource for certain time. | The customer should shift or curtail home resource for certain time. | The customer should shift or curtail home resource for certain time. |

Table 3.2: Different demand response mechanism

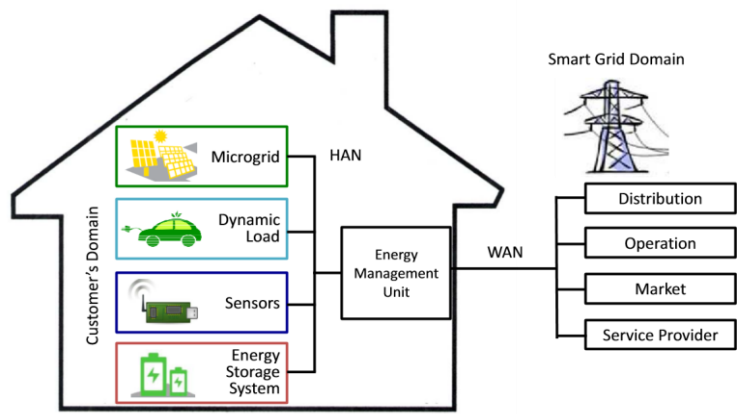

Figure 3.7: Architecture of DSM framework

## 3.5 Home energy management system

Associated with the demand response action and local microgrid installation of many residential customers, there are many researchers proposing the idea of a zero-energy building or a smart greenhouse assisted by a home energy management system (HEMS). To some extent, a HEMS is actually a part of the smart grid on the consumption side, which analyzes the microscopic parameters of appliances. It is employed to collect data from home appliances (e.g. solar panel, electric vehicle, geo thermal, LED lamp) using smart meters and pervasive sensors, and then to optimize power demand and supply based on this collected local information. A typical HEMS usually focuses on power consumption monitoring and standby power reduction. With the increasing demand for intelligent and personalized services, the so-called context-aware systems have been implemented in a smart home to support these personalized services with machine learning and reasoning mechanisms. These innovative designed systems have advantages that can offer adaptive energy service prediction with respect to the power consumption patterns and the related human activities. According to the users' activities and predefined requirements, HEMS along with context-aware systems can reason through the adaptive DR actions by analyzing incentives and power management policies [37].

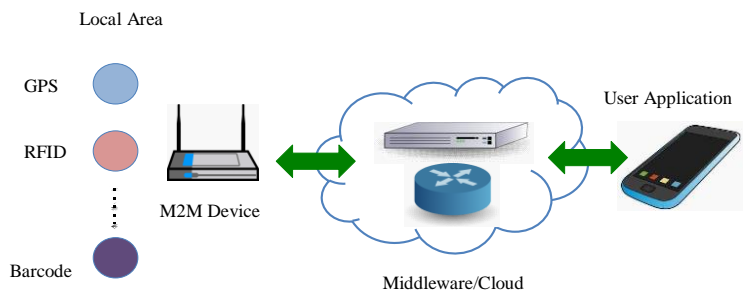

Figure 3.8: M2M communication in HEMS

In recent years, HEMS gradually combines context-aware systems to improve energy consumption efficiency and resident satisfaction and comfortable level. However, since conventional HEMS need excessive resource consumption and long-term pattern analysis to have energy consumption pattern generation, these systems usually have a lot of time delay and energy mismatch. In contrast, modern HEMS usually exploits various embedded sensors (e.g. smart meters or intelligent monitoring sensors) and advanced ICT infrastructure (e.g. cloud computing platform or fog computing platform) to support managing complex applications and services. Large-scale usage of embedded sensors and IoT technology will lead to a great rise in machine-to-machine (M2M) communications over wired and wireless links, which also requires enormous computing resources (Figure 3.8).

On the other hand, along with the ability to control home appliances intelligently and efficiently using HEMS, integrating local microgrids with renewable energy systems becomes increasingly important. Compared with the past applications that are usually limited to heating water or heating a room through single energy source, the current applications also include remotely operating home appliances or adjusting lighting system, as well as linking the existing appliances directly with renewable energy generation systems and energy storage systems (ESSs). The applicability of various renewable energy resources is continuously increasing in recent years. Future zero energy buildings or smart green homes (which produce and provide the electricity themselves through local microgrid without the external supply of electricity, maximizing energy efficiency) could be realized by context-awareness technology and the M2M technique based intelligent home energy management system. More importantly, through the utilization of two-way communication means, it is possible to use home area network (HAN) to connect different smart devices and measurement units to an energy management system (EMS) and manage the operation of electric appliances in an economic way. With a focus on residential EMSs, a large number of research and demonstration projects have been done recently and related findings have been published in different scientific papers [39, 40, 41]. As an example, authors of [39] proposed a residential energy system to provide grid support services and manage different distributed energy resources (DERs), considering the minimum operation cost. Authors of [41] developed the energy scheduling strategy and domestic energy management for a residential building with taking technical and operational issues into account. An innovative single-objective energy management algorithm for domestic load scheduling has been outlined in [40] with the similar goal to minimize energy consumption cost. Other authors have also investigated such residential energy management problem in multiple ways with taking into account a time-domain simulation, incentive-based DR actions and price-elastic load shifting [42]. As can be observed from the related literature, there is a large and growing body of research addressing the home energy management problem within smart residential micro-grids considering different objectives and related constraints. Although some literature try to cover the extent of these problems of home energy scheduling in future smart grids with networked microgrid, several challenges are still associated with intelligent energy management production and energy consumption units. In [43], the author proposes a multi-objective dispatching model of a residential smart EMS (MOEMS) in order to coordinate different DERs and smart household devices. Demand response is generally performed in the residential district through HEMS, since residential districts will be aware of and more sensitive to the electricity price with shiftable, controllable, flexible, interruptible, deferrable, elastic and dispatchable appliances, e.g., PHEV, washer, and dryer. For these appliances, users are only concerned about the results of whether their tasks are finished within a certain time period without referring to the particular intermediate steps. This fact implies that their aggregate energy consumption should not be less than a threshold before a particular deadline. Based on two-way communications in HEMS, smart metering or advanced metering infrastructure (AMI) could gather detailed information regarding users' electricity usage patterns and provide automatic control to household appliances, which forms the core functionality of HEMS. As illustrated in Figure 3, there is an energy consumption scheduler (ECS) embedded in the smart meter at each household, whose role is to control the ON/OFF switch and operating mode of each appliance. The dynamic electricity price signal can be obtained from the power utility and the user's energy demand exchanging information via HAN. The smart meter acts as a controller that coordinates all appliances to satisfy the user's requests. After the demand response, the smart meter will send ON/OFF or UP/DOWN control commands with specified operating modes to all appliances, according to the optimized energy consumption schedule [10].

The electricity price and local energy demand are exchanged via the HAN in the household. On the other hand, the communication between the household and the power utility is based on WAN. In this way, the smart meter or AMI is able to automatically coordinate all electric devices via ON/OFF control commands with specified operating modes.

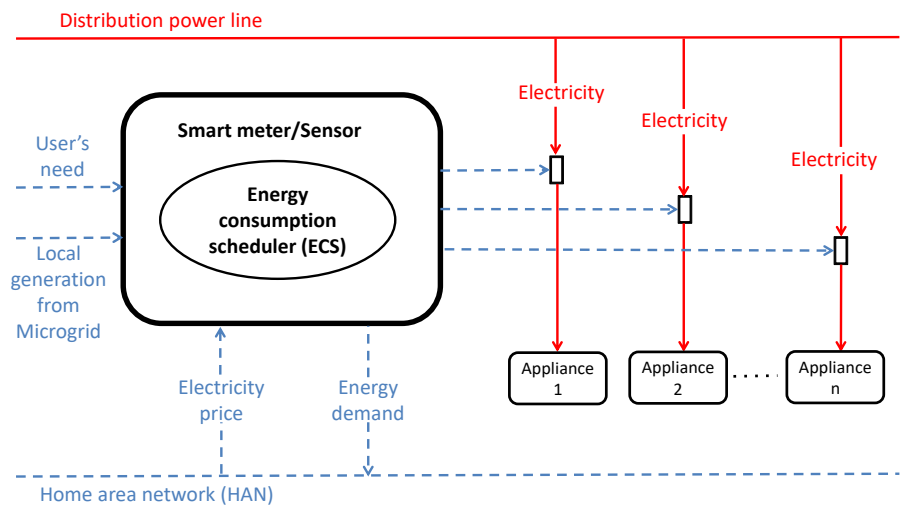

Figure 3.9: Home energy management system

## 3.6 Energy management with SCADA

Supervisory control and data acquisition (SCADA) is a computer system for gathering and analyzing the real-time data of power system. SCADA systems have been widely used to monitor and control a plant or equipment in industries since the 1980s. The supervision and management of a microgrid with SCADA are shown in Figure 3.10 and [44], based on SCADA systems through web services. A human machine interface (HMI) is developed with the objective of facilitating the interaction between users and systems.

In order to manage the two-type microgrid, namely in island mode and connected mode, two different architectures are needed and proposed in [44]. In the hierarchical architecture there is a central controller element that governs the local controllers of the energy sources. In the decentralized architecture, the control is replicated in the local controllers of the energy sources. Nowadays, the utilization of a SCADA makes configuration and supervision of microgrids much easier than conventional field-test. Any layman can easily make necessary modifications to introduce new energy sources in the entire system or setting the scheduled response sensitivity. The end-users can adapt the system by simply receiving useful information, such as the cost of fuel or the efficiency curve, by means of intuitive graphical user interface (GUI) and simple screens, which the user can even personalize. The user can access total knowledge of the functioning of every element to supervise the system at all times, and therefore to make decision to resolve unforeseen situations quickly. The system also allows users to control different functions such as starting or stopping sources of energy in real-time. As illustrated in Figure 3.10, this SCADA has a series of layers that give the user access to different field variables. The open platform communications (OPC) client accesses the OPC server and can obtain the variables from the general controller. The information is sent to the XML-DA server, which allows REST (Representational State Transfer) requests, making it possible to visualize the information using a web interface [44]. In addition, along with the increasing interactions between customers who have local microgrids and the capability to provide energy for the neighborhood community, an S-SCADA (Social Supervisory Control and Data Acquisition) concept is presented in some literatures [45]. This approach that consists of various computational tools is capable of linking the physical world with the social or cyber world in order to support community development. The traditional SCADA system require expertise to gather and analyze real time data; professional interfaces (e.g. HMI) are needed for system operators, who are expected to have technical knowledge of the plant that is being controlled. The S-SCADA is designed to create dummy interfaces that can easily present processed information to users who have no technical knowledge of the system.

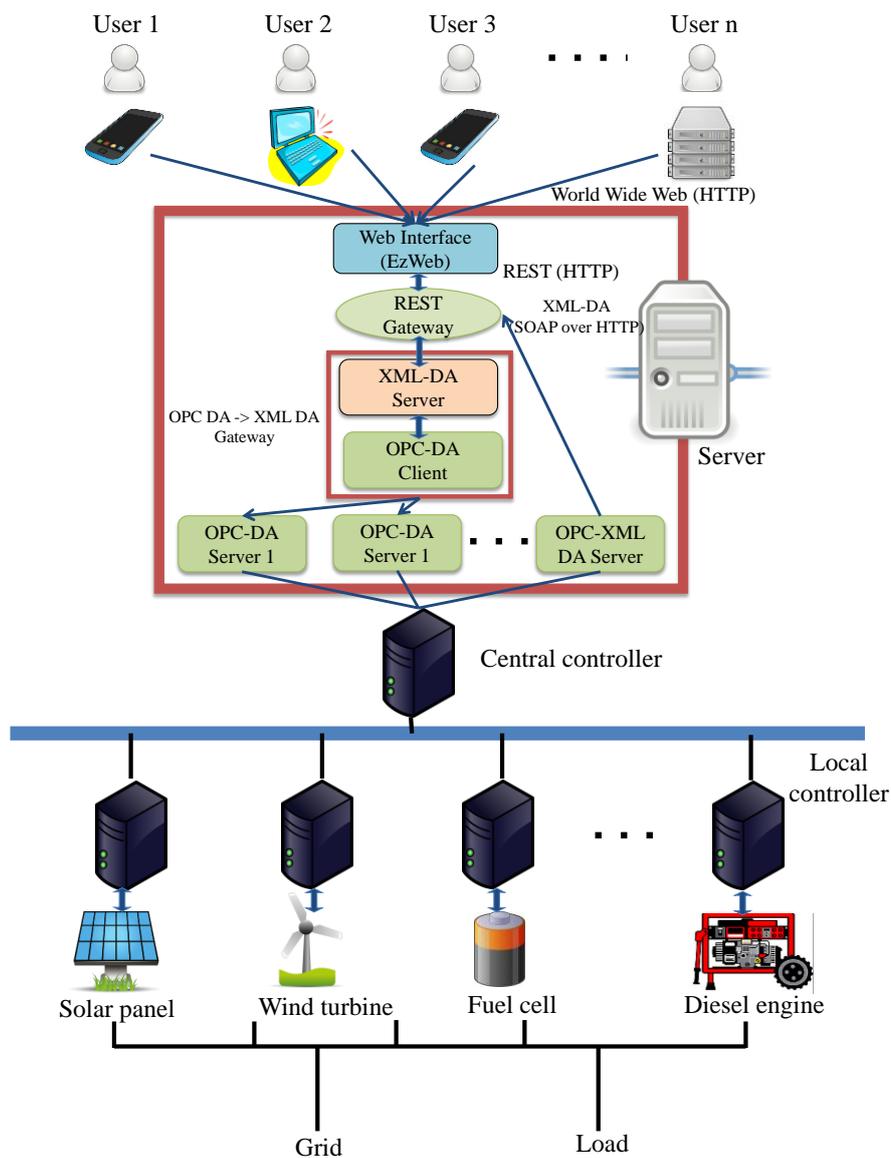

Figure 3.10: General system architecture of SCADA for microgrid management

## 3.7 Supporting infrastructure

The improvement of power equipment technologies has been a great contribution to the development of smart grids. In this section, we mainly focus on smart meter measurements, which are considered as the evolution of the existing grid structure. Besides that, with the advancement of computer communication technologies, smart meters can likely enhance the operation efficiency and reliability of power systems.

### 3.7.1 Smart Meters Systems

The smart meter is widely used as a smart grid technology. It is an electronic device for recording electric energy consumption. It also obtains information and communicates it back to the utility for monitoring and billing. The initial implementation of the smart meter technology was in the fields of commercial and industrial because most customers have the need for more sophisticated rates and more granular billing data requirements. For over 15 years, electronic meters have been used effectively by utilities in delivering accurate billing data for at least a portion of their customer bases[46]. The previous technology for collecting energy consumption data from meters is called automated meter reading (AMR). It provides only one-way communication, from the home to the utility. Evolved from the foundations of AMR, the advanced metering infrastructure (AMI) was developed in around 2005. The AMI technology differs from the traditional AMR by providing two-way communication between meters and the central system. The development from the AMR to the AMI, and their functionalities, are summarized in Figure 3.11.

In essence, the smart grid is an innovative reconstruction from the aspects of transmission and distribution and the smart meter system is an important integrant of the smart grid infrastructure in data collection and communication. In essence, there are three components in the smart meter system: the advanced metering device, communication network management, and data processing system.

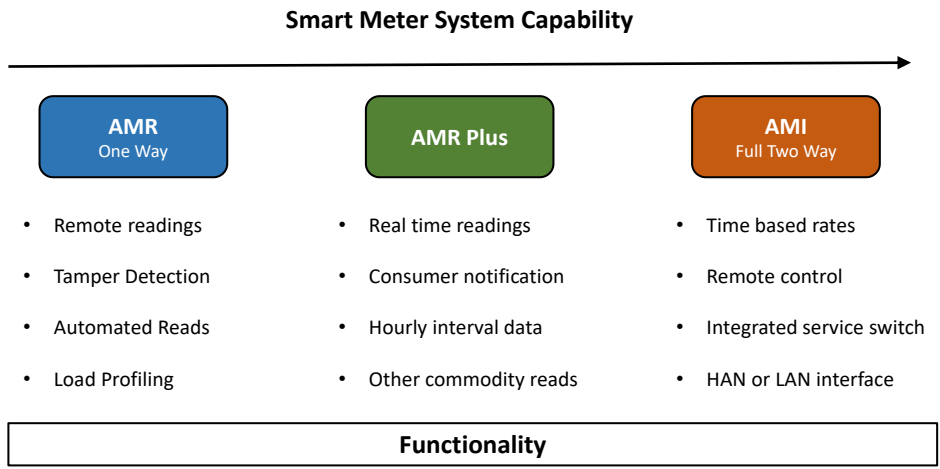

Figure 3.11: Smart Meter Technology Development

Smart meters operate on a two-way communication, so an internal memory component is needed. Smart meters also allow consumers to track their own energy use on the Internet and/or with third-party computer programs. The two-way nature of smart meter systems allows for sending commands to operate grid infrastructure devices, such as distribution switches and recloses, to provide a more reliable energy delivery system. This is known as distribution automation [46]. Smart meter systems have many benefits, no matter who the consumer or electrical company is. From the consumers' point of view, the smart meter enables the delivery of a rapid report to the central system when a tampering happens. This can effectively help reduce the rate of theft and improve security. Besides that, everyday billing information is available for every customer so that each one can manage his/her own usage of appliances and, consequently, lower their bill. From the electrical companies' point of view, the management of power consumption data from every meter can be easily gathered and processed and naturally, the procedure of billing can be made fast with the help of the two-way communication.

### 3.7.2 Advanced Metering Infrastructure (AMI)

AMI is developed from AMR. It is a technology that provides a connection between system operators and consumers. On one hand, the information is available for consumers so they are more aware of and can make adjustments to energy usage. On the other hand, system operators can improve the service and billing process based on the data provided by AMI. The AMI infrastructure consists of home network systems, including thermostats and other in-home controls, smart meters, communication networks from the meters to local data concentrators, back-haul communications networks to corporate data centers, meter data management systems and finally, data integration into existing and new software application platforms [47]. Figure 3.12 illustrates the overview of the AMI system and how it connects with in-home controls and communication networks.

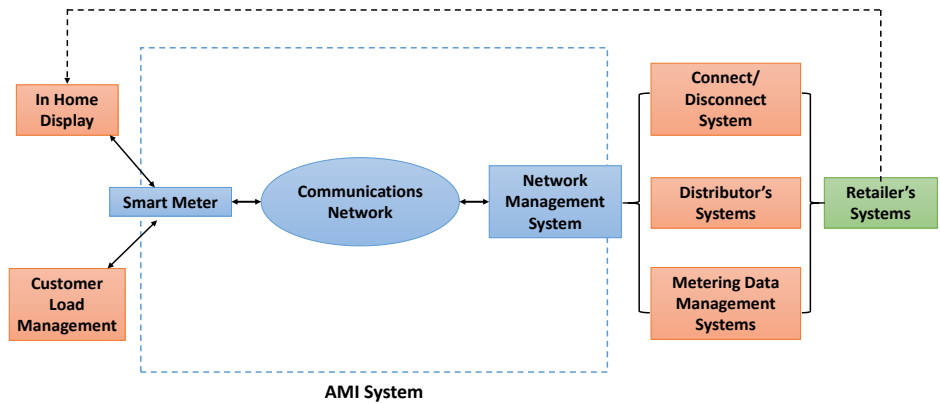

Figure 3.12: Overview of AMI

## Smart Meters

For the residents, meters simply record the total energy consumption over a period of time. However, there are also other functions, such as power quality monitoring, net metering, and load limiting, as well. This helps emissions and carbon reductions and eventually improves energy efficiency. Meters show this information to every customer and are likely to cause a reduction of energy usage.

## Communication Infrastructure

The communication infrastructure in AMI builds a platform between consumers, the utility, and the electrical load. In terms of security, the infrastructure must employ bi-directional communication standards. Local concentrators that used to collect data from different meters are commonly implemented. They also transmit the data to the central server, the bandwidth of which should be considered, based on consumer services and other requirements.

## Home Area Networks (HAN)

HANs provide every consumer with a portal that connects smart meters to electrical devices. They also act as the consumer's agent. The consumer can check their energy usage and its cost through the in-home display and set limits for the utility to control the loads from wasting energy.

### 3.7.3 Privacy and Security of Smart Meters

Cyber security is regarded as one of the biggest challenges in smart grids. Vulnerabilities may allow an attacker to penetrate a system, obtain private user information, gain access to control of the software, and alter load conditions to destabilize the grid in unpredictable ways. We must realize that the advanced infrastructure implemented in smart grids, on one hand, empowers us with more powerful mechanisms to defend against attacks from outside, but on the other hand, exposes many new vulnerabilities [48]. In this part, we mainly discuss some security and privacy issues due to the deployment of smart meters.

#### 3.7.3.1 Security in Smart Metering

Attacks on smart meters can be classified as physical (external tampering, neutral bypass, missing neutral, etc.), electrical (over/under voltage, circuit probing, etc.), and software and data [49]. Smart meters raise several serious security issues:

1. The risk of widespread fraud: when smart meters are widely implemented, meter readings could be manipulated. The industry is concerned with the reliability of the returned read data. This could ruin the service provider's reputation and lead to unpredictable losses;

2. Excessive technical regulation: equipment suppliers argue that equipment costs have been pushed up and that there is nearly no benefit of continuing to move forward with smart grids. This attitude can be harmful to the prospect of fixing security problems;

3. Strategic vulnerability: a remote off switch exists in all electricity meters. A potential adversary could switch off devices by using unpredictable cyberattacks;

4. Conflict of interest: while it is the governments that prefer to cut energy use, in most countries, the meters will be controlled by energy retailers, whose goal is to maximize sales. Meanwhile, the competition authorities should worry about whether giving energy retailers vast amounts of data on customers will adversely impact competition via increased lock-in [50];

5. Lack of universal standards: communication between meters and appliances is important. An authorized standard can improve the interoperability and management in the central system, eventually relieving the competition. However, many countries cannot even decide on the architecture for connecting appliances to meters.

#### 3.7.3.2 Privacy in Smart Metering

Smart meters also have potential problems for customer privacy. Retailers have the ability to obtain huge amount of data from meters or other electric devices. Not only would this reveal energy usage information, but personal habits, behaviors and preferences could also be disclosed to some interested parties. Smart meter data, which consists of granular, fine-grained, high-frequency type energy usage measurements, can be used by others either maliciously or inadvertently using existing or developing technology to infer types of activities or occupancies of a home for specific periods of time. Analysis of granular smart meter energy data may result in [51]:

1. Invasion of privacy and intrusion of solitude;

2. Near real-time surveillance;

3. Behavior profiling;

4. Endangering the physical security of life, family and property;

5. Unwanted publicity and embarrassment (e.g., public disclosure of private facts or the publication of facts which place a person in a false light).

6. Determine how many people are home and at what times;

7. Determine what appliances you use when, e.g., washer, dryer, toaster, furnace, A/C, microwave, medical devices ... the list is almost endless, depending on the granularity of the data;

8. Determine when a home is vacant (for planning a burglary), who has high priced appliances, and who has a security system;

9. Law enforcement can obtain information to identify suspicious or illegal behavior or later determine whether you were home on the night of an alleged crime;

10. Landlords can spy on tenants through an online utility account portal;

11. For consumers with plug-in electric vehicles, charging data can be used to identify travel routines and history;

12. Utilities can promote targeted energy management services and products;

13. Marketers could obtain information for targeted advertising.

In order to address the privacy problems related to smart meters, some approaches have been discussed or proposed:

1. Compress the meter readings and use random sequences in the compressed sensing to enhance the privacy and integrity of the meter readings [52];

2. For billing or performing calculations, improve the existing protocol and ensure the data is accurate without disclosing it;

3. Encrypt all consumption data, make most personal information anonymized, and collect and directly send both of them to the central system.

## 3.8   Conclusion and Future Research Trends

As a cutting-edge technology, Microgrids feature intelligent energy management systems and sophisticated control, and will dramatically change our energy infrastructure. The modern microgrids are a relatively recent development with high potential to bring distributed generation, distributed energy storage devices, controllable loads, communication infrastructure, and many new technologies into the mainstream. As a more controllable and intelligent entity, a microgrid has more growth potential than ever before. However, there are still many open questions, such as the future business models and economics. What's the cost-benefit to the end-user? How should we systematically evaluate the potential benefits and costs of control and energy management in a microgrid?

[25] Y. Guo, J. Xiong, S. Xu, and W. Su, "Two-Stage Economic Operation of Microgrid-Like Electric Vehicle Parking Deck," 2015.

[26] Y. He, B. Venkatesh, and L. Guan, "Optimal scheduling for charging and discharging of electric vehicles," *IEEE Transactions on Smart Grid*, vol. 3, no. 3, pp. 1095–1105, 2012.

[27] T. Senjyu, R. Kuninaka, N. Urasaki, H. Fujita, T. Funabashi, M. Ieee, and S. Member, "Power system stabilization based on robust centralized and decentralized controllers," *2005 International Power Engineering Conference*, pp. 905– 910 Vol. 2, 2005.

[28] R. Olfati-Saber, J. A. Fax, and R. M. Murray, "Consensus and Cooperation in Networked Multi-Agent Systems," *Proceedings of the IEEE*, vol. 95, pp. 215– 233, jan 2007.

[29] Wei Ren, R. Beard, and E. Atkins, "A survey of consensus problems in multiagent coordination," in *Proceedings of the 2005, American Control Conference, 2005.*, pp. 1859–1864, IEEE, 2005.

[30] N. Rahbari-Asr, U. Ojha, Z. Zhang, and M.-y. Chow, "Incremental Welfare Consensus Algorithm for Cooperative Distributed Generation/Demand Response in Smart Grid," *IEEE Transactions on Smart Grid*, vol. 5, pp. 2836–2845, nov 2014.

[31] R. Mudumbai, S. Dasgupta, and B. B. Cho, "Distributed Control for Optimal Economic Dispatch of a Network of Heterogeneous Power Generators," *IEEE Transactions on Power Systems*, vol. 27, pp. 1750–1760, nov 2012.

[32] Z. Zhang and M. Y. Chow, "Incremental cost consensus algorithm in a smart grid environment," in *IEEE Power and Energy Society General Meeting*, 2011.

[33] V. Loia and A. Vaccaro, "Decentralized economic dispatch in smart grids by selforganizing dynamic agents," *IEEE Transactions on Systems, Man, and Cybernetics: Systems*, vol. 44, no. 4, pp. 397–408, 2014.

[34] H. K. Nunna, A. M. Saklani, A. Sesetti, S. Battula, S. Doolla, and D. Srinivasan, "Multi-agent based Demand Response management system for combined operation of smart microgrids," *Sustainable Energy, Grids and Networks*, vol. 6, pp. 25–34, jun 2016.

[35] H. T. Haider, O. H. See, and W. Elmenreich, "A review of residential demand response of smart grid," *Renewable and Sustainable Energy Reviews*, vol. 59, pp. 166–178, jun 2016.

[36] E. Karfopoulos, L. Tena, A. Torres, P. Salas, J. G. Jorda, A. Dimeas, and N. Hatziargyriou, "A multi-agent system providing demand response services from